\documentclass[aip,jcp,reprint,twocolumn]{revtex4-2}
\usepackage{graphicx}
\usepackage{dcolumn}
\usepackage{bm}
\usepackage{graphicx}
\usepackage{dcolumn}
\usepackage{bm}
\usepackage{dcolumn}
\usepackage{amsmath}
\usepackage{graphicx}
\usepackage{dcolumn}
\usepackage{bm}
\usepackage{multirow}
\usepackage{xcolor}
\usepackage{easyReview} 

\newcommand{\ro}  { {\bf r}}
\usepackage[caption=false]{subfig}
\usepackage[normalem]{ulem}
\begin{document}
	\title{Effect of Counterion Size on Polyelectrolyte Conformations and Thermodynamics}
	\author{Souradeep Ghosh}
	\affiliation{Department of Physical Sciences and Centre for Advanced Functional Materials, Indian Institute of Science Education and Research
		Kolkata, Mohanpur 741246, India}
	\author{Arindam Kundagrami}
	\affiliation{Department of Physical Sciences and Centre for Advanced Functional Materials, Indian Institute of Science Education and Research
		Kolkata, Mohanpur 741246, India}
	
	\begin{abstract}
		We present a theoretical model to study the effect of counterion size on the effective charge, size, and thermodynamic behavior of a single, isolated, and flexible polyelectrolyte (PE) chain. We analyze how altering counterion size modifies the energy and entropy contributions to the system, including the ion-pair free energy, excluded volume interactions, entropy of free and condensed ions, and dipolar attraction among monomer-counterion pairs, which result in competing effects challenging intuitive predictions. The PE self energy is calculated using Edwards-Muthukumar Hamiltonian, considering a Gaussian monomer distribution for the PE. The condensed ions are assumed confined within a cylindrical volume around the PE backbone. The dipolar and excluded volume interactions are described by the second and third virial coefficients. Assumption of freely-rotating dipoles results in a first-order coil-globule transition of the PE chain. A more realistic weaker dipolar attraction, parameterized in our theory, shifts it to a second-order continuous transition. We calculate the size scaling-exponent of the PE and find exponents according to the relative dominance of the electrostatic, excluded volume, or dipolar effects. We further identify the entropy- and energy-driven regimes of the effective charge and conformation of the PE, highlighting the interplay of free ion entropy and ion-pair energy with varying electrostatic strengths. The crossover strength, dependent on the counterion size, indicates that diminishing sizes favor counterion condensation at the expense of free ion entropy. The predictions of the model are consistent with trends in simulations, and generalize findings of the point-like counterion theories.
	\end{abstract}
	
	\maketitle
	\section{INTRODUCTION}
		
	The conformational behavior of flexible uncharged polymers in different solvents is well-understood. In general, good solvents lead to extended conformations and bad solvents result in collapsed globules. For flexible polyelectrolytes (PE), however, in the presence of counterions, conformations may undergo a coil-globule transition, irrespective of the solvent type.\cite{Oosawa1971-nv,Khokhlov1982,Brilliantov1998,Solis2000,Muthukumar2004,DOBRYNIN2005,Dua2005,Kundagrami2008,Kundagrami2010,Chi2013,Kundu2014,Tom2016,Muthukumar2017,Tom2017,Mitra2017,Muthukumar2023} This transition depends on the extent of counterion adsorption, influenced by the interplay of the electrostatic energy for the formation of monomer-counterion bound ion-pairs and the translational entropy of the free counterions and salt ions. The monomer-counterion bound pairs come at the cost of the tranlational entropy of the free ions, and results in a train of dipoles along the chain backbone. Solvent quality, aided by the short-range dipolar attraction, leads to the coil-globule transition in a poor solvent. In a good solvent, however, the transition is caused by the latter. It has been shown that larger (bulky) counterions\cite{Gavrilov2016, Gordievskaya2018-SM} or surfactant-like counterions\cite{vonFerber2003} can prevent the coil-globule transition. Such counterion specificity also plays a role in controlling bulk properties such as viscosity and conductivity of PE systems\cite{Klooster1983-1, Klooster1983-2,Kim1992,Liberatore2010,Mori2014,Rumyantsev2016}. However, the influence of the counterion size on the interaction between the localized ions and the PE and on the equilibrium behavior of the PE and the counterions has largely been absent in early theoretical models,\cite{Khokhlov1982,Brilliantov1998,Solis2000,Muthukumar2004,DOBRYNIN2005,Dua2005,Kundagrami2010,Kundagrami2008,Kundu2014} even though its importance was recognized earlier.\cite{Oosawa1971-nv} Early computer simulations typically accounted for counterion specificity through finite size, fixed at equal to or smaller than a monomer.\cite{Winkler1998,Liu2002,Liu2003} The effect of the counterion nature on the conformations of a single flexible PE and related thermodynamics has only recently been explored theoretically\cite{Tom2016,Tom2017}, despite a wealth of experimental and molecular simulation data being available for decades on the subject.\cite{Klooster1983-1,Klooster1983-2,Kim1992,Yasumoto2006,Wyatt2010,Malikova2012,Mori2014,Malikova2015,Rajeev2017,Bodrova2007,Hua2012,Philippova2013,Rumyantsev2016,Gavrilov2016,Gordievskaya2018-SM,Gordievskaya2018,Tom2016,Tom2017,Kos2020} The effect of counterion size on single polyelectrolyte molecules has in detail been studied using molecular dynamics simulations.\cite{Gavrilov2016,Gordievskaya2018-SM,Gordievskaya2018,Tom2016,Tom2017} Additionally, there have been theoretical models focusing on the influence of counterion size in PE gels.\cite{Bodrova2007,Hua2012,Philippova2013,Rumyantsev2016}
	
	The swelling behavior of polyelectrolyte gels depends on the size and type of counterions, which affect the ion association, the counterion condensation process, and osmotic pressure of `free' counterions within the gel.\cite{Satoh2000,Bodrova2007,Hua2012,Philippova2013,Rumyantsev2016} The volume transition theory of PE gels with charge regularization includes a variable dielectric mismatch parameter that implicitly accounted for the salt ion diameter.\cite{Hua2012} Reentrant swelling with intermediate counterion size are found to be caused by dipolar attraction and excluded volume interactions.\cite{Philippova2013} Small counterions cause gel collapse, while large counterions prevent it, by suppressing ion pairing and increasing swelling.\cite{Bodrova2007,Philippova2013,Rumyantsev2016} The solvent specificity of PE gel collapse has also been observed in experiments.\cite{Satoh2000} The viscosity and conductivity of PE gels are also reported to be dependent upon ion size, where small ions condense more to decrease free ion concentration, reducing conductivity.\cite{Rumyantsev2016}
	
	In polymer solutions, the bulk viscosity is influenced by the conformation of the polymer chains. The presence of counterions of specific sizes and solvation characteristics affect the PE conformations, and in turn the viscosity. For example, in entangled PE solutions (Xanthan gum), larger salt counterions, both monovalent and divalent, lead to higher viscosities.\cite{Liberatore2010} Studies on other polymer solutions, such as PSS and grafted PAA, have shown that viscosity is nearly proportional to the hydrodynamic size of the counterions.\cite{Mori2014, Kim1992} The choice of solvent also impacts the viscosity behavior of polyelectrolyte solutions in the presence of salt. For instance, in a PAA solution in methanol, Li$^+$ was found to induce higher viscosity compared to Na$^+$.\cite{Klooster1983-1, Klooster1983-2} The drastic drop in viscosity for Na$^+$ suggests a collapse driven by dipolar interactions.\cite{Kundu2014} Moreover, in aqueous methanol, the molar conductivity was found inversely proportional to the hydrodynamic size of the counterions, indicating loosely bound hydrated ions\cite{Klooster1983-2}. However, at higher methanol concentrations, the trend reverses, suggesting strengthened ion-pair formations with decreased solvation effects. The conductivity in water is found to be higher compared to methanol due to a lesser condensation or association of counterions in water that has a high dilectric constant.\cite{Klooster1983-1, Klooster1983-2}
	
	For strongly charged polyelectrolytes in dilute solutions, simulations\cite{Gavrilov2016, Gordievskaya2018-SM, Gordievskaya2018} have demonstrated that the size of counterions and the strength of Coulomb interaction significantly influence the conformational behavior of the PE chains. Bulky counterions lead to swollen conformations of PE chains, where the counterions are loosely bound to the chain backbone and move freely around the PE and in the solution.\cite{Gavrilov2016, Gordievskaya2018-SM} The conformational behavior of a dipolar polymer chain was suggested to be influenced by the interplay between electrostatic and excluded volume interactions,\cite{Gordievskaya2018} which can potentially be controlled in experiments by varying the solvent composition or temperature.
	
		
	Regarding the counterion distribution near the chain backbone, neutron scattering has shown that a compact double layer around the ionene backbone is formed when Br$ ^- $ ions are present, in contrast to F$ ^- $ counterions\cite{Malikova2012}. Mixtures of counterions induce a selectivity in condensation of small ions against large ions, resulting in intra-polymer micro-phase separation and core-shell micro-structure formation within polyelectrolyte globules, observed in simulations.\cite{Gavrilov2016, Gordievskaya2018-SM}. Additionally, in PE brushes, bridging interactions due to monovalent ions have been reported\cite{Das2021}. The smaller ions, such as Li$ ^+ $, bridge more strongly than the larger ions, such as Cs$ ^+ $.

    These detailed studies on polyelectrolyte systems with finite-size counterions, as compared to point-like counterions, reveal their richness and motivate further investigation. A few previous works\cite{Winkler1998,Brilliantov1998,Liu2003,Muthukumar2004,Cherstvy2010,Kundu2014} have theorized the PE chain collapse due to dipolar attraction, even considering the orientational restrictions of the dipoles.\cite{Cherstvy2010} However, these studies did not consider the finite size effects or excluded volume interactions of the counterions, which recent simulations and experiments have suggested to play a significant role.\cite{Hua2012,Philippova2013,Rumyantsev2016,Gavrilov2016, Gordievskaya2018-SM, Gordievskaya2018} Specifically, a virial expansion model\cite{Tom2017} focused on regimes of Coulomb strengths (or Bjerrum length, $\ell_{B}$) leading to collapsed conformations, and pointed out that inclusion of only till the third virial term is sufficient for such purposes.\cite{Tom2016} However, the thermodynamic aspects resulting from condensation of finite size counterions, and related conformational transitions, for the entire range of the Coulomb strength have not been looked into yet, to the best of our knowledge. A theory that incorporates explicit calculations of the PE self-energy from an interaction Hamiltonian, accounts for dipolar interactions and also finite size effects of counterions through the virial coefficients to the lowest order, and remains applicable for all physically accessible $\ell_{B}$ values, shows potential for a more comprehensive understanding of the system.

    To this end, we aim to build a general, minimalistic theoretical model for a single, isolated, and flexible PE chain with finite-size counterions, investigating the effect of the counterion size on the PE's effective charge, size, and thermodynamics. Our analytical model focuses on the size variation of counterions, which results in modifications to the energy and and entropy components of the system including the ion-pair energy, excluded volume interactions, volume entropy of free ions, volume entropy of condensed ions assumed confined to a cylindrical volume conformal to the PE backbone, dipolar interactions captured through the second virial coefficient, and the third virial coefficient required to stabilize the collapse of the chain. The increase in counterion size reduces the gain in free energy due to both ion-pair formation and the volume entropy of free ions. As a result, it leads to non-monotonic thermodynamic effects making intuitive predictions challenging. To construct the theory, we use the Edwards-Muthukumar interaction Hamiltonian,\cite{Edwards1979,Muthukumar1987,Muthukumar2004} which captures the self-energy of the PE chain through segment–segment electrostatic and excluded volume interactions, including dipolar interactions, and also the conformational entropy of the PE chain. The derived generic free energy is extremized through a Gaussian trial Hamiltonian, following Flory.\cite{Flory1950}

	The use of freely rotating dipolar-pair interactions and short-range repulsions through the second and third virial coefficients in the free energy obviates the need of any new parameter in the theory, in addition to the three major ones - the Bjerrum length, Debye screening length, and dilectric mismatch parameter. The dipolar interactions lead to a first-order coil-globule transition of the PE at reasonably high Coulomb strengths, which shifts to a continuous and second order transition with increasing counterion size, that opposes the chain collapse progressively. The parameterization of the dipolar interaction, assuming an over-estimation from the use of freely rotating dipoles, also shifts the transition to be second order. In addition, we also calculate the size scaling exponents, and observe a variety of scaling behavior as an interplay of electrostatic (monopolar), excluded volume, and dipolar interactions. We further derive the thermodynamics by identifying the enthalpy- and entropy-driven regimes, as functions of Coulomb strength and counterion size, both in the absence and presence of a moderate salt concentration.

	\section{Theory}
	
	For the theoretical model, we consider a linear and flexible polyelectrolyte (PE) chain composed of $N$ identical ionizable groups as repeat units or monomers, each of diameter $\ell$ carrying a monovalent negative charge, with counterions of finite size (of diameter $ r_{c} $), in a dilute solution with volume $\Omega$. With total $N$ counterions, the degree of counterion condensation $\alpha=M/N$ is defined as the ratio of the number of charge-compensated monomers (the monomer on which a counterion has condensed), $M$, to the total number of monomers in the chain. Degree of ionization of the chain is defined as ($ f=1-\alpha $).
	
	The number density of the externally added monovalent salt, which is assumed to fully dissociate into $n_+$ cations and $n_-$ anions, is given by $c_s=n_+/\Omega \equiv n_-/\Omega$.  The dimensionless monomer density in the solution, denoted as $\bar{\rho}$, can be expressed as $\bar{\rho} = N/(\Omega / r_c^3)$. Similarly, we define $\bar{c}_s = n_{s} /(\Omega / r_c^3)$, where $n_s=n_+=n_-$.
	
	The radius of gyration ($R_g$) characterizes the size of the PE chain. The free energy ($F$) of the system comprises the self-energy - conformal, electrostatic, and excluded volume - of the chain, and the entropic and enthalpic contributions of the condensed and mobile counterions. The free energy depends on two independent variables, namely $M$ (or, equilvalently, $\alpha$ or $f$) and $R_g$. The objective of the theory\cite{Muthukumar2004,Kundagrami2010} is to self-consistently evaluate the equilibrium values of $M$ and $R_g$ by minimizing the free energy $F$, which is also a function of the electrostatic and other parameters (say, $N, \Omega, \ell_B, n_s, r_c$ etc.), with respect to these variables, and to find the effect of counterion size ($r_c$) on the thermodynamic and conformational properties. 
	
	The free-energy of the polymer chain is formulated by using the Edwards-Muthukumar Hamiltonian\cite{Edwards1979,Muthukumar1987,Muthukumar2004,Muthukumar2023}, and applying required additions and modifications as mentioned below. The variational free energy of the system obtained from such Hamiltonian and free ion contributions allows the size effects of counterions to manifest in larger scales, through the size and overall charge of the PE chain, the analysis of which is the main aim of this work. The total free energy is obtained from the following contributions.
	
\vskip 0.25cm

\begin{figure}[!htbp]%
		\centering
		\includegraphics[width=0.9\linewidth]{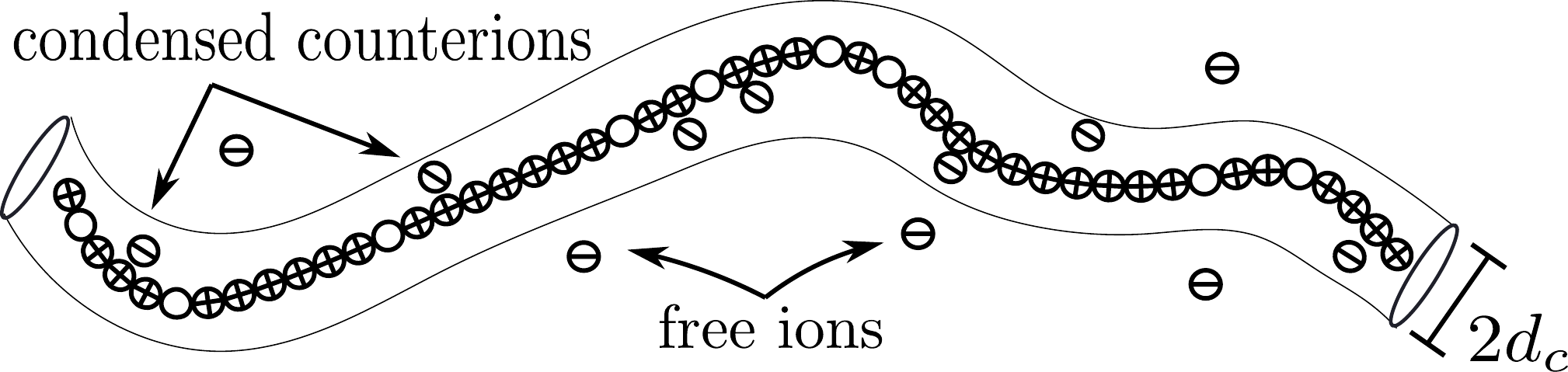}
		\includegraphics[width=0.4\linewidth]{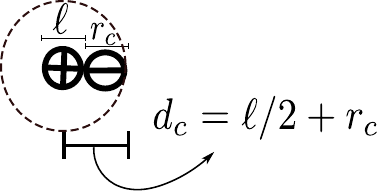}
		\caption{\textbf{Cylindrical volume for condensed counterions around the PE chain:} To calculate the entropy of condensded counterions and electrostatic free energy of ion-pairs, motivated by evidence from simulations that the counterion density sharply peaks near the chain backbone\cite{Hofmann2001,Liu2003}, we consider the counterions inside a hypothetical cylindrical volume of radius $d_{c}=\ell/2+r_c$ around the chain contour. }
		\label{fig:si-schematic}
	\end{figure}

\noindent {\bf A. Entropy of condensed counterions:} It is often assumed that there are $\binom{N}{M}$ ways to distribute $M$ counterions over $N$ monomers of a PE chain. The condensed counterions are, however, mobile along the chain contour, and monomer-counterion pairs typically do not form frozen dipoles, but show thermal fluctuations\cite{Barrat1996}. Therefore, to account for the entropy of the condensed counterions,it is reasonable to consider a volume, $\Omega_{c}$, for which the outer boundary is a cylinder of radius $d_c=\ell⁄2+r_c$, and the inner boundary is set by the monomer length ($\ell$), conformal with the chain backbone (Fig. \ref{fig:si-schematic}). Within this volume, $M$ counterions are randomly adsorbed (condensed) along the chain backbone. The translational volume entropy of such condensed counterions confined to a volume of ${\Omega}_c$ is given by
\begin{align}
S_1=k_B \log \left(\frac{\bar{\Omega}_c !}{(\bar{\Omega}_c-M) ! M !}\right),
\end{align}
leading to the free energy $F_1=-TS_1$ given by 
	\begin{align}\label{F1}
		\frac{F_{1}}{k_B T}=\bar{\Omega}_c\left[\left(1-\frac{M}{\bar{\Omega}_c}\right) \log \left(1-\frac{M}{\bar{\Omega}_c}\right)+\left(\frac{M}{\bar{\Omega}_c}\right) \log \left(\frac{M}{\bar{\Omega}_c}\right)\right],
	\end{align}	
	where, $\bar{\Omega}_c \equiv \Omega_c/r_c^3 = N\left[\left(\pi (0.5 +\tilde{r}_{c})^2-1\right)\right]/\tilde{r}_c^3$, and $\tilde{r}_{c}=r_{c}/\ell$ is the dimensionless diameter of the counterions.

\vskip 0.25cm	

\noindent {\bf B. Entropy of free ions:} The free ion entropy associated with $ (N-M+n_{+}) $ number of the uncondensed counterions and	the salt cations and $ n_{-} $ number of coions (i.e., the free mobile ions in the solution) in the volume $ \Omega $ is $ k_{B} \log({{\Omega}^{N-M+n_{+}}}/{(N-M+n_{+})! n_{-}!}) $, and $ F_{2} $, the free energy due to such entropy, is given by
	
	\begin{equation}\label{F2}
		\frac{F_{2}}{k_B T}=N\left[\left(f+\frac{\bar{c}_s}{\bar{\rho}}\right) \log \left(f \bar{\rho}+\bar{c}_s\right)+\frac{\bar{c}_s}{\bar{\rho}} \log \bar{c}_s-\left(f+\frac{2 \bar{c}_s}{\bar{\rho}}\right)\right].
	\end{equation}
	 The dimensionless monomer density in the solution is given by, $\bar{\rho}=N/(\Omega / r_c^3)=(N \ell^{3}/\Omega)(r_c^{3}/\ell^{3})=\tilde{\rho} \tilde{r}_{c}^3$ and $ \bar{c}_s=n_{s} /(\Omega / r_c^3)=(n_s \ell^{3}/\Omega)(r_c^{3}/\ell^{3})=\tilde{c}_s \tilde{r}_{c}^3 $, where $\tilde{\rho}=\rho \ell^3$ and $\tilde{c}_s=c_s \ell^3$.
\vskip 0.25cm	
	
\noindent{\bf C. Free energy of ion density fluctuation:} In the limit of low salt, that is $\kappa \ell \rightarrow 0$, the Helmholtz free energy due to counterion density fluctuations approaches\cite{Mcqurrie, Muthukumar1996}
	\begin{align}
		\frac{F_{3}}{k_{B}T}=-\frac{\Omega \kappa^3}{12 \pi}=\frac{N\sqrt{4\pi}\widetilde{\ell}_{B}^{3/2}}{3\bar{\rho}}(f\bar{\rho}+2 \bar{c}_{s})^{3/2},
		\label{DH}
	\end{align}
where the free ions considered for the expression is the same ones as in $F_2$. Here, $\widetilde{\kappa}=\sqrt{ 4 \pi \widetilde{ \ell}_{B}(f\bar{\rho}+2 \bar{c_{s}})/\tilde{r}_{c}^3}$ is the dimensionless inverse of Debye screening length, and $\widetilde{\kappa}=\kappa \ell$.  With finite size of the counterions, and when the size becomes significant, comparable, or even bigger than the monomers, Eq. \ref{DH}, which is only a limiting result for $\kappa r_c \rightarrow 0$, ideally needs to be replaced by the full expression\cite{Mcqurrie,Muthukumar2002} of the free energy given by 
	\begin{align}
		\frac{F_{3}}{\bar{\Omega} k_{\mathrm{B}} T}=-\frac{1}{4 \pi}\left[\log (1+\widetilde{\kappa} \tilde{r}_{c})-\widetilde{\kappa} \tilde{r}_{c}+\frac{1}{2}(\widetilde{\kappa} \tilde{r}_{c})^2\right],\label{F3-mcqurrie}
	\end{align}
where the finiteness of the counterions is accounted for by taking them as spheres of diameter $r_c$. 

However, for low salt, $\kappa r_c \ll 1$, even if the counterion size becomes large. In this work, we have taken very low amounts of salts for a few results, and even with $r_c/\ell=4$, Eq. \eqref{DH} remains sufficient. For an analysis with high salt and large counterions, Eq. \eqref{F3-mcqurrie} needs to be used. 
	
\vskip 0.25cm	
	
\noindent{\bf D. Free energy of ion-pair formation:} The accumulation of oppositely charged counterions near the PE chain can be characterized by both counterion condensation and a localized ionic atmosphere, and both result in qualitatively similar thermodynamic effects \cite{Shen2017}. The free energy contribution from the electrostatic 
attraction between the charged monomers and `condensed' counterions can be calculated exactly, if we have the knowledge of the bound counterion density profile. However, for our analysis, we make the assumption that such profile, or the related pair correlation function, is sharply peaked near the chain backbone\cite{Hofmann2001,Liu2003,Chen2022}, and the monomers and respective counterions form dipoles with the shortest possible dipole length\cite{Khokhlov1994, Winkler1998, Muthukumar2004,Cherstvy2010}. The gain in free energy due to the formation of an ion pair	associated with the adsorption of one counterion to a charged segment is, therefore, $ -e^2 /(4 \pi \epsilon_0 \epsilon_{l} d_{mc} )$, where $ \epsilon_{l} $ is the local dielectric constant and $d_{mc}$ is the dipole length between the charge of the monomer and the counterion.
	
	The adsorption free energy gain due to $ M $ number of counterion-monomer pairs is then given by
	\begin{equation}
		\frac{F_{4}}{k_{B}T}=-(1-f)N\delta \tilde{\ell}_{B},
		\label{F4}
	\end{equation}
	where $ \delta=(\epsilon \ell/\epsilon_{l} d_{mc}) $, and $\widetilde{d}_{mc}=d_{mc}/\ell=(\ell+r_{c}) /2\ell$. The presence of a local binding constant is modeled by the phenomenological parameter $\delta$,\cite{Khokhlov1994,Kramarenko2002,Muthukumar2004,Kudlay2004,Kundagrami2010,
Mitra2023,Ghosh2023} which in a coarse-grained way qualitatively captures the drop in the local dielectric constant close to the organic PE (or protein) chain bakcbone, compared to its bulk value in a polar solvent. $\delta$ as a parameter (although inadequately, in the absence of a microscopic theory\cite{Li2013,Chen2022-PNAS,Muthukumar2023}) addresses the fact that the local dielectric environment is significantly different for PEs and proteins from that of isolated small ions\cite{Li2013,Dinpajooh2016, Fischermeier2016,Chowdhury2019}, the effect recognized in early investigations\cite{Mehler1984, Lamm1997, Rouzina1998}. The limited accessibility and disorientation of polar solvent dipoles close to to chain backbone result in a continuous rate of increase of $\epsilon_{l}$ with distance from the chain backbone,\cite{Mehler1984} but in this model, for simplicity, a single value of $\epsilon_{l}$, lower than the bulk value $\epsilon$, is taken as a parameter. The electrostatic interaction of two like-charged counterions condensing on two adjacent monomers is accounted for, to some extent, by the dipolar interaction, that is discussed later.

\vskip 0.25cm	
	
\noindent{\bf E. Self energy of the PE chain:} In terms of a general Hamiltonian $H$ that comprises the potentials for the monomer-monomer interactions of the PE chain, the free energy $F_5$ of the chain originating from such Hamiltonian will be given by
	\begin{align}\label{Partition_function}
		e^{-\beta F_5}=\int D \mathbf{R}\left(s\right) \exp (-\beta H),
	\end{align}
	where $\beta=1 / k_B T$. The integral $D \mathbf{R}\left(s\right)$ is a conformational integral of the canonical partition function of the polyion. $\mathbf{R}\left(s\right)$ is the position vector of the chain at the arc length variable $s\left(0 \leq s \leq N\right)$, where $N$ is the number of monomers in the chain. The interaction Hamiltonian $H$, developed by Edwards and Singh\cite{Edwards1979}, and extended by Muthukumar for charged systems\cite{Muthukumar1987,Muthukumar2004}, can be expressed in terms of the following interactions between monomers: a) their connectivity ($ H_{0} $), b) the short range interactions among monomers or condensed counterions ($ H_{ex} $), whereas the monomers may be separated by a large distance (long-range) along the contour of the PE, and the interactions include repulsive, non-electrostatic excluded volume interactions or attractive dipolar interactions, and c) the screened repulsive electrostatic interaction between charge uncompensated monomers ($ H_{el} $). Hence 
	\begin{align}\label{eq:H}
		&{H}=H_{0}+H_{ex}+H_{el},
	\end{align}
	where the components ($H_{0},~H_{ex}\text{ and }H_{el}$) are given by
		\begin{align}
			&H_{0}=\frac{3}{2 \ell^2}\int_{0}^{N} d s\left(\frac{\partial {\bf R}\left(s\right)}{\partial s}\right)^{2}\label{eq:H_0}\\
			&H_{ex}=\omega\ell^{3} \int_{0}^{N} ds \int_{0}^{N} ds^\prime \delta( \mathbf{R}(s)-\mathbf{R}(s^\prime)),\label{eq:H_ex}\text{ and }\\
			&H_{el}=\frac{f^2\ell_{B}}{2}\int_{0}^{N} ds \int_{0}^{N} ds^\prime    \frac{\exp \left(-\kappa \left|{\bf R}(s)-{\bf R}(s^\prime)\right|\right)}{\left|{\bf R}(s)-{\bf R}(s^\prime)\right|}.\label{eq:H_el}
		\end{align}
	 The arguments of the $\delta$-functions in the above integrals (Eq. \ref{eq:H_ex}) denote the difference in contour vectors corresponding to the monomer pair involved in the short range interactions (excluded volume or dipolar), where $w$ is the interaction strength. The interactions among charge uncompensated monomers are governed by a screened Coulomb electrostatic potential, with the screening parameter $\kappa$. The counterion condensation has been addressed at the mean-field level, resulting in the coefficient $f^2$ in Eq. \ref{eq:H_el}.
	
	Directly evaluating the partition sum using the aforementioned Hamiltonian (Eqs. \ref{eq:H}, \ref{eq:H_0}, \ref{eq:H_ex}, and \ref{eq:H_el}) can be a rather intricate task. Instead, a variational procedure\cite{Muthukumar1987}, that involves a trial Hamiltonian achieved by redefining the Hamiltonian of Eq. \ref{eq:H} as 
	\begin{align}
		H=&H_{\text{trial}}+(H-H_{\text{trial}}),
	\end{align}
	where
	\begin{align}\label{trial-Hamiltonian}
		H_{\text{trial}}=\frac{3}{2 \ell\ell_{1}}\int_{0}^{N} d s\left(\frac{\partial {\bf R}\left(s\right)}{\partial s}\right)^{2},
	\end{align}
can be employed.
	Here, $\ell_{1}$ represents the variational parameter that characterizes the effective expansion factor of the polyion in comparison to its Gaussian size\cite{Edwards1979,Muthukumar1987,Muthukumar2004,Muthukumar2023}.
	The mean-field assumption is based on the (Gibbs-Bogoliubov) inequality,
	\begin{align}
		\left\langle\mathrm{e}^{-\beta H}\right\rangle_{H_{trial}} \geq \mathrm{e}^{-\beta\left\langle H\right\rangle_{H_{trial}}},
	\end{align}
	which implies that the free energy,
	\begin{align} \label{F5-trial}
		\widetilde{F}_5=\langle \beta(H_0-H_{trial})\rangle_{H_{trial}}+\langle \beta H_{ex}\rangle_{H_{trial}}
		+\langle \beta H_{el}\rangle_{H_{trial}},
	\end{align}
	needs to be extremized with respect to the charge ($ f $) and size (expansion factor, $ \ell_{1} $) of the  polyelectrolyte . If one shifts to the polymer coordinate (the spatial coordinate ${\bf r}$) the Hamiltonian can be approximately recast in terms of the monomer density profile of the PE chain\cite{Flory1950,podgornik1993,Muthukumar2012,Ghosh2023}. If one assumes a spherically symmetric Gaussian distribution, the monomer density centered at $\mathbf{r}_{0}$ and positioned at $\mathbf{r}$ can be expressed as
	\begin{align}\label{rho-def}
		&\rho_{n0}(\ro)=N\left(\frac{3}{4 \pi R_{g}^{2}}\right)^{3 / 2} \exp \left[-\frac{3 (\left|\ro-\ro_{0}\right|)^{2}}{2 R_{g}^{2}}\right].
	\end{align}
	Under the assumption of uniform expansion of the PE chains\cite{Edwards1979,Muthukumar1987,Muthukumar2004,Muthukumar2023}, the average dimensionless radius of gyration of the chain can be obtained as
	\begin{align}\label{rg_def}
		\widetilde{R}_{g}=\sqrt{\frac{N \tilde{\ell}_{1}}{6}},
	\end{align}
	where $\widetilde{R}_{g}=R_g/\ell$ and $\tilde{\ell}_{1}=\ell_{1}/\ell$.			
	
	Using the Fourier transform (in ${\bf k}$-space) of the monomer density profile (Eq. \ref{rho-def}) and integrating the averaged interaction of monomers (Eq. \ref{F5-trial}), the total free energy contribution due to the polymer degrees of freedom included in the Hamiltonian (Eq. \ref{eq:H}) can be obtained in the form
	\begin{align}
		\frac{F_{5}}{k_{B} T}&=F_{51}+F_{52}+F_{53}\nonumber\\
		& =\frac{3}{2}\left[\widetilde{\ell}_{1}-1-\log \widetilde{\ell}_{1}\right]+ \left(\frac{9}{2 \pi}\right)^{3 / 2}  \frac{w \sqrt{N}}{\widetilde{\ell}_{1}^{3 / 2}} \nonumber\\
		&+\frac{f^2  N^{2}  \widetilde{\ell}_{B}}{2} \Theta_{s}\left(a\right),\label{F5}
	\end{align}
	where,
	\begin{align}
		\Theta_{s}\left(a\right)=\frac{2}{\pi}\left[\sqrt{\frac{\pi\widetilde{\kappa}^{2}}{4a}}-\frac{\widetilde{\kappa} \pi}{2} \exp{\left(a\right)} \text{erfc}\left(\sqrt{a}\right)\right],
	\end{align}
	and $ a={\widetilde{\kappa}^{2} \tilde{R}_{g}^2}/{3} = \widetilde{\kappa}^{2} N \widetilde{\ell}_{1}/18$.
	
	The effective two-body interaction parameter $w$ (Eq. \ref{F5}) is the most important quantity in this work. We note that the short-range $\delta$-function interactions (Eq. \ref{eq:H_ex}) can be of several type - the repulsive excluded volume interaction, attractive charge-dipole, and dipole-dipole interactions etc.. The short-range attractive interactions, in this case involving the dipoles, effectively modify the excluded volume parameter\cite{Pincus1998, Muthukumar2004, Kundu2014, Mitra2023}. The size and number of dipoles formed on the chain backbone play a critical role in determining the strength of such interactions and, in turn, in the equilibrium behavior of polyelectrolytes. Considering the counterion adsorption, one determines that $(1-f)N$ out of $N$ monomers are paired with counterions, and the remaining $fN$ monomers are charge uncompensated. The overall excluded volume parameter in the mean-field can thus be written as,\cite{Muthukumar1996,Pincus1998}
	\begin{align}\label{excluded-volume}
		w=f^2 w_{m m}+(1-f)^2 w_{dd}+f(1-f)w_{md},
	\end{align}
	where $w_{mm}$, $w_{dd}$, and $ w_{md} $ are the strengths of the short-range two-body interactions arising from, respectively, the usual, non-electrostatic excluded volume interaction between uncompensated monomers, electrostatic attraction between a pair of ion-pairs (a dipole pair), and the electrostatic attraction between an uncompensated monomer and an ion-pair (a monopole-dipole pair). The limiting cases are as follows. In the absence of any condensed counterion, $f=1$, and $w=w_{mm}$. Conversely, when all the counterions are condensed, $f=0$, and $w=w_{dd}$. We may note that for an extended chain $w$, which is a two-body short-range interaction parameter, is not effective. Therefore, we may assume the first term containing $w_{mm}$ negligible compared to the second term in our analysis. 
	
	In this model it turns out that $w_{md}$ for the monopole-dipole interaction consists of coefficients and have dependency on the electrostatic parameters ($\widetilde{\ell}_B$ and $\delta$) similar to that of $w_{dd}$ for the dipole-dipole interaction, and both are attractive. In addition, for a first order collapse of the chain a significant amount of counterion condensation occurs ($f \rightarrow 0$), leading to the dipole-dipole pair interaction being dominant over the monopole-dipole interaction. Hence, in the subsequent calculations, we ignore the effects arising from monopole-dipole interactions and the last term in Eq. \ref{excluded-volume}.
	
	As discussed before, the short-range attractive interaction between dipoles embedded on the chain backbone can be represented by a $\delta$-function potential\cite{Muthukumar2004, Cherstvy2010, Mitra2023,Kundu2014} with the two-body strength parameter as $w_{dd}$. $w_{dd}$ can be calculated the usual way using the Mayer's function\cite{Pathria} once the actual interaction potential is known. Considering the dielectric mismatch near the chain backbone, and assuming that counterions may adsorb at random directions perpendicular to the local chain axis (freely rotating dipoles), and the chain being flexible, the interaction energy $U_{dd}(r)$ between a pair of dipoles separated by a distance of $r$ can be expressed by
	\begin{align}
	\label{udd}
		\frac{U_{dd}(r)}{k_B T}= \begin{cases}+\infty & \text { if } r \leq \sigma \\ -\left(4\pi /3\right)\left(\widetilde{d}_{mc}^3\delta\widetilde{\ell}_{B}/\widetilde{r}^3\right)^2 & \text { if } r>\sigma\end{cases}
	\end{align}
	where $\widetilde{d}_{mc}=d_{mc} / \ell$, $ \widetilde{r}=r/\ell $, and $ \sigma $ corresponds to the hard sphere contact distance. In addition, when the counterions are of a similar size to the monomers or larger, the short-range repulsion (usual non-electrostatic excluded volume interaction) in the dipole-pair interaction needs to be considered as well. Putting back the potential, ${U_{dd}(r)}/{k_B T}$, into the Mayer function, we get,
	\begin{align}
		f(r) = \begin{cases}-1, & r \leq \sigma \\ -\beta U_{dd}(r), &  r \geq \sigma.\end{cases}
	\end{align}
	Hence, the second virial coefficient (in units of volume) can be obtained as,
	\begin{align}
		w_{dd}^\prime& =-\int_0^{\infty} 4 \pi r^2 f(r) d r \nonumber\\
		& =-\int_0^\sigma 4 \pi(-1) r^2 d r-\int_\sigma^{\infty} 4 \pi(-\beta U_{dd}(r)) r^2 d r.
	\end{align}
	Assuming $\sigma \simeq d_{mc}$, the dimensionless strength parameter of dipole-dipole interactions that contributes to the the excluded volume interaction can be obtained as,
	\begin{align}
	\label{finalwdd}
		w_{dd}=\frac{w_{dd}^\prime}{\ell^3} \equiv\frac{4 \pi  \widetilde{d}_{mc}^3}{3}-\frac{16}{9} \pi ^2 \widetilde{d}_{mc}^{3} (\delta\widetilde{\ell}_{B})^2.
	\end{align}
	The first term arises from repulsive interactions between pairs of dipoles (for $r \leq \sigma$, Eq. \ref{udd}). Such excluded volume contributions increase with the increasing number of bound counterions (Eq.
 \ref{excluded-volume}), as pointed out in earlier literature.\cite{Oosawa1971-nv} Furthermore, since a polymer chain cannot be more compact than a sphere with $ R_g\sim N^{1/3} $, to ensure a physically realistic result for a collapsed chain, one needs to consider the three-body interaction through the third virial coefficient,\cite{Muthukumar1984,Muthukumar1989,Kundagrami2010,Kundu2014,Wang2017} denoted by $ w_3 $. We include it in our calculation with the additional free energy term given by
	\begin{align}
		\frac{F_{6}}{k_B T}=\frac{w_3}{\widetilde{ \ell}_1^3}.
		\label{F6}
	\end{align}
	Instead of taking $ w_3 $ as a parameter, as was previously done,\cite{Muthukumar2004,Dua2005,Kundagrami2010,Kundu2014,Mitra2017} we calculate it explicitly with the equation,
	\begin{align}
		w_3=-\frac{1}{3 \ell^6} \int_0^{\infty} \int_0^{\infty} f_{12} f_{13} f_{23} d^3 r_{12} d^3 r_{13}, 
	\end{align}
	where, in general notation, $ f_{ij} $ and $ r_{ij} $ are the Mayer function and the distance between particle $ i $ and $ j $, respectively. To evaluate this integral, we first fix the positions of particles $ 1 $ and $ 2 $ (such that $r_{12}<d_{mc}$ ) and let particle $ 3 $ take all possible positions so that we can effectively integrate over the variable $r_{13}$.\cite{Pathria} To achieve an analytical form, we assume hard sphere potential in our case, with the hard sphere contact being the dipole length $d_{mc}$. The third virial coefficient will then become
	\begin{align}
		w_3= \frac{5 \pi^2 d_{mc}^6}{18 \ell^6} \equiv \frac{5 \pi^2 \tilde{d}_{mc}^6}{18}.\label{w3}
	\end{align}
	
	As constructed, the total free energy, $F=\sum_i F_i$, $i=1$ to 6,  depends on two variables: $\widetilde{\ell}_{1}$, which represents the effective expansion factor of the mean square end-to-end distance of the PE chain compared to its Gaussian size, and the degree of ionization $f$ of the PE chain.  
	
	It should be noted that the free energy described above is applicable only for a single polyelectrolyte (PE) chain in a dilute solution. It remains valid for all degrees of ionization or ionizability of the PE, as well as all temperatures. However, it is only applicable for salt concentrations that are not too high, such that $ \kappa^{-1} \ge \ell_{B} $ or $ c_s \le (8\pi\ell_{B}^3 )^{-1} $ for a monovalent salt.

	\section{Results and Discussion}

	The system consists of a solution containing one polyelectrolyte chain with finite-size counterions and also small molecular salt. Several interactions such as Coulomb energy of ion-pairs, screened Coulomb repulsion among charge-uncompensated monomers, density fluctuations of the mobile ions in the solution, excluded volume interactions among monomers as well as counterions, and dipolar attractions between monomer-counterion ion-pairs are present, and have been described by the free energy components in our model [Eqs. \eqref{F1}, \eqref{F2}, \eqref{DH}, \eqref{F4}, \eqref{F5}, and \eqref{F6}]. The equilibrium total free energy $F$ (expressed as $\sum_{i=1}^{6}F_i$) is determined by a self-consistent minimization with respect to the size, given by the effective expansion factor of the PE chain, $\widetilde{ \ell}_1$, and the degree of counterion condensation, $\alpha$ (or, equivalently, degree of ionization, $f=1-\alpha$). The most important parameter of this study is the counterion size, $\widetilde{r}_c$. The temperature, $T$, and the bulk dielectric constant, $\epsilon$, in terms of the dimensionless Bjerrum length, $\widetilde{\ell}_{B}$, the degree of polymerization, $N$, the monomer density, $\bar{\rho}$, the monovalent salt density, $\bar{c}_{s}$, and the dielectric mismatch parameter, $\delta$, (a function of both $r_c$ and the local dilectric constant $\epsilon_l$) are the other parameters of the problem. We set the monomer density by fixing the dimensionless volume of the system, $\Omega/\ell^3=2\mathrm{x}10^6$, where $\ell$ represents the size of a monomer. For a chain of length $N=1000$, this results in the dimensionless monomer density, $\bar{\rho}=0.0005~\tilde{r}_c^3$. Both monomers and counterions are taken monovalent. 
		
	Our primary focus is on the effect of counterion specificity, through its size, on the equilibrium configurations of the PE chain, degree of counterion condensation, the size scaling exponents, and system thermodynamics, through evaluation of individual free energy components. 
	
	\subsection{Effect of local dielectric constant ($ \epsilon_{l} $) on the conformational behavior of the PE chain}
	
	We first benchmark the general problem of counterion condensation by the known results for a fully ionizable PE, taking the counterion size equal to the monomer size as is traditionally done\cite{Oosawa1971-nv,Khokhlov1982,Winkler1998,Pincus1998,Brilliantov1998,Liu2002,Liu2003,Muthukumar2004,Kundagrami2010,Chi2013}. There are two major differences in the formulation of our model compared to the previous ones. First, a volume entropy instead of combinatorial entropy for the condensed counterions, confined to a cylinder conformal to the chain backbone, is considered (Fig. \ref{fig:si-schematic} and Eq. \ref{F1}) and second, electrostatic self-energy of the PE chain has been calculated differently, in a simpler way (Eqs. \ref{Partition_function} to \ref{F5}). Here we briefly note the key results, obtained by the minimization of the total free energy with respect to the thermodynamic variables size and charge, $\widetilde{ \ell}_1$ and $f$, respectively, of the PE chain for a set of values of the local dielectric constant ($\epsilon_{l}$), represented by $\delta$. Importantly, to benchmark and compare with the previous results of charge interactions, the excluded volume and dipolar interactions are ignored for the time being, and no additional salt is taken ($w=0.0$, $ w_3=0.0 $, and $\bar{c}_s=0.0$). 
	
	\begin{figure}[!htbp]%
		\centering
		\includegraphics[width=1\linewidth]{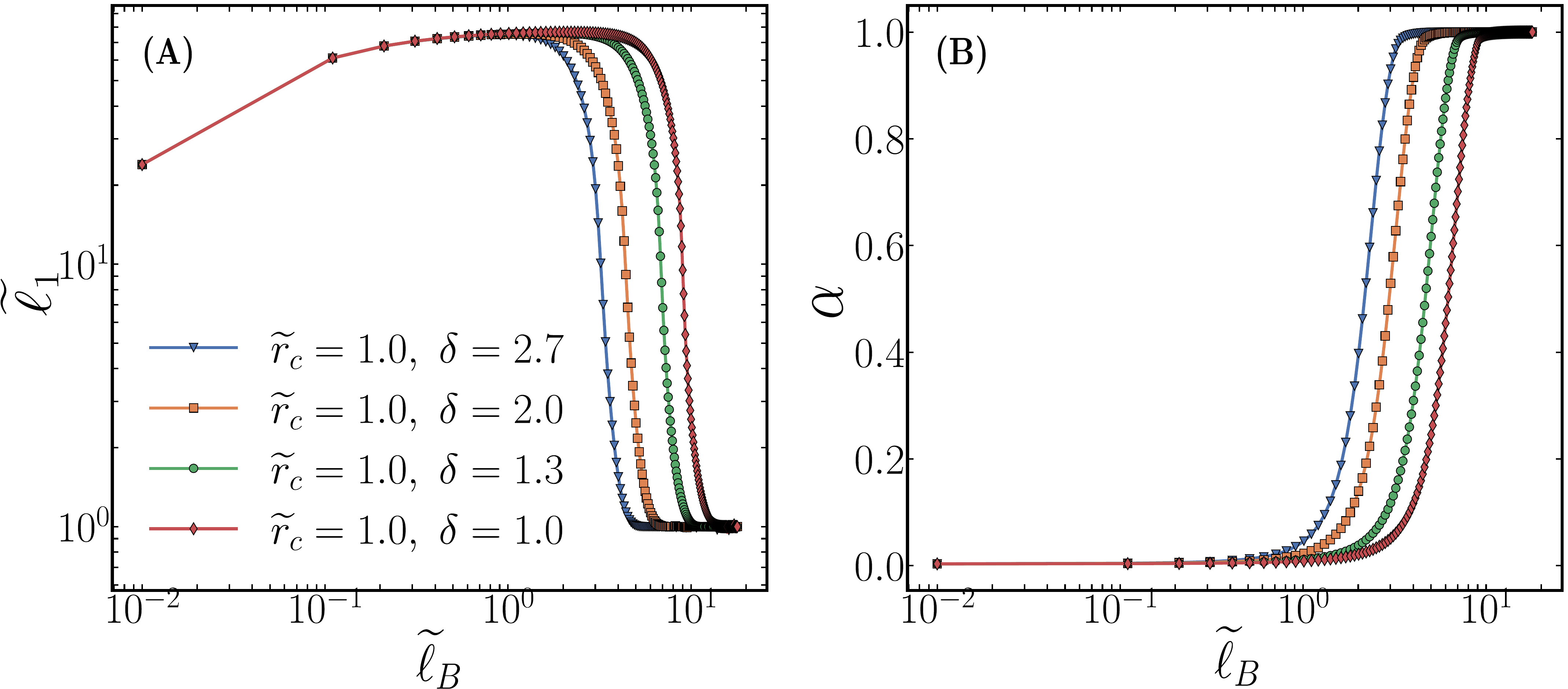}
		\caption{\textbf{Size and charge of a single, isolated, flexible PE chain (ignoring excluded volume and dipolar interactions):} (A) The size $(\widetilde{\ell}_1)$ and  (B) the degree of counterion condensation $(\alpha=1-f)$  of the PE chain are plotted as functions of the Bjerrum length $\widetilde{\ell}_B$ for different local dielectric constants ($\delta=\epsilon/\epsilon_l$ with counterions having the same size of monomers, i.e., $\widetilde{r}_c=1.0$). Excluded volume and dipolar interactions are ignored ($w=w_3=0.0$). The PE charge and size decrease significantly with a decreasing local dielectric constant, $\epsilon_{l}$. The other parameters are: $N=1000,~\bar{c}_s=0.0$, and $\bar{\rho}=0.0005~\tilde{r}_c^3$.}
		\label{fig:spm-l1-f-vs-lbt-fixed}
	\end{figure}

	The degree of counterion condensation ($\alpha=1-f$) and the size of the PE chain ($\widetilde{\ell}_{1}=6\widetilde{R}{g}^2/N$) are obtained as a function of the Bjerrum length ($\widetilde{\ell}_{B}$), as shown in Fig. \ref{fig:spm-l1-f-vs-lbt-fixed}(A-B). As expected, in the weak electrostatic regime (low $\widetilde{\ell}_{B}$ or high temperatures - note that the bulk dielectric constant does not affect the product $\delta \widetilde{\ell}_{B}$), the counterion  adsorption is minimal ($\alpha\sim 0$, $f \sim 1$), and the thermalized chain is Gaussian with $\widetilde{\ell}_{1}\sim 1$. As the Coulombic effect strengthens (higher $\widetilde{\ell}_{B}$ or lower temperature), first the electrostatic repulsion among charged monomers expands the chain. With further increasing $\widetilde{\ell}_{B}$, counterions start to adsorb onto the chain backbone, reducing such repulsion, that leads to deswelling of the chain. At very high $\widetilde{\ell}_{B}$ values, all the counterions get adsorbed ($\alpha\sim 1$, $f \sim 0$), and in the absence of electrostatic repulsion the chain assumes Gaussian configuration once more ($\widetilde{\ell}_{1} \sim 1$), provided that the excluded volume interaction and dipolar attraction are both ignored. Therefore, $\widetilde{\ell}_{1}$ exhibits non-monotonic variation with $\widetilde{\ell}_{B}$, while $\alpha$ increases monotonically. These results are qualitatively very similar to Ref.\cite{Muthukumar2004}, but in our model the free energy of the PE chain is calculated by assuming a Gaussian segment distribution, resulting in a simpler self-energy expression (Eq. \ref{F5}). 
	
	The variation in local dielectricity ($\epsilon_l$) affects $\alpha$, and in turn $\widetilde{\ell}_{1}$. By considering the closest distance between the condensed counterion and monomer allowed by a hard sphere contact, the expression for $d_{mc}$ simplifies to $d_{mc} = \ell/2 + r_c/2 \equiv \ell$. As the counterions have the same size of the monomers in this case ($r_c = \ell$), we can further simplify to $\delta = \epsilon \ell / \epsilon_l d_{mc} = \epsilon / \epsilon_l$.

	Furthermore, for a fixed $\tilde{\ell}_B (\geq 1)$, a lower value of the local dielectric constant ($\epsilon_{l}$) leads to a greater accumulation of counterions near the chain backbone [Fig. \ref{fig:spm-l1-f-vs-lbt-fixed}(B)], due to a higher electrostatic energy gain, quantified by, $- \delta\widetilde{\ell}_B=-e^2/4\pi\epsilon_{0}\epsilon_{l}k_{B}T$. In essence, the degree of ionization is highly sensitive to the dielectric mismatch $\delta$, analogous to $\widetilde{\ell}_B$. \cite{Khokhlov1994,Kramarenko2002,Muthukumar2004,Kudlay2004,Kundagrami2010,	Mitra2023,Ghosh2023,Muthukumar2023}

	\subsection{Counterions with finite size}
	
	In this section, we include the effective two-body interaction parameter (Eq. \ref{excluded-volume}), and consider the effect of counterion size on counterion adsorption, chain conformations, and thermodynamics. As the counterions can have large sizes, higher than the monomers, one must consider the excluded volume interactions among them. More importantly, the adsorption of counterions forms dipoles on the PE chain, the interactions among which (as described in the second and third terms of Eq. \ref{F5}) will have significant effect on the conformations and thermodynamics of the chain. 

    With an increasing size of the counterions, both the Coulomb free energy gain of counterion-monomer pairs ($F_4$, Eq. \ref{F4}) and free ion entropy ($F_2$, Eq. \ref{F2}) decrease. These two competing and nonlinear thermodynamic contributions ($F_2$ and $F_4$) effectively set the degree of counterion condensation, which in turn dictates the size of the PE chain. Therefore, in general it is hard to predict the trends with changing counterion size, just on the physical or intuitive grounds. Given the modest set of parameters we have used, however, it is apparent that with decreasing counterion size the electrostatic gain in ion-pair free energy ($F_4$) wins over the loss in entropy, due to the loss of a freely roaming counterion to counterion-monomer pair formation ($F_2$). This we find violated in a few cases, as we shall see later. Furthermore, the counterion size affects the length of the monomer-counterion dipole. The pairwise dipolar interaction is captured through the second virial coefficient $w_{dd}$ (Eq. \ref{excluded-volume}), whereas the three-body interaction is incorporated via the third virial coefficient $w_3$ (Eq. \ref{F6}), the latter being required to provide stability against collapse, as detailed in Eq. \ref{w3}. It is notable that the introduction of $w_{dd}$ and $w_3$ above, in their current form (Eq. \ref{F5} and \ref{F6}), does not lead to any new parameter to the analysis. 	 
	
	After such introduction of the second and third virial coefficients, we continue the minimization of the free energy with respect to size and charge, $\widetilde{\ell}_{1}$ and $\alpha$ (or $f$), respectively. As we vary the counterion size in this section, $\epsilon/\epsilon_{l}=2$ is kept constant, equivalent of taking the same pair of PE backbone and the polar solvent, for this part of the analysis with variable counterion size. 
	
		\begin{figure}[!htbp]%
		\centering
		\includegraphics[width=1\linewidth]{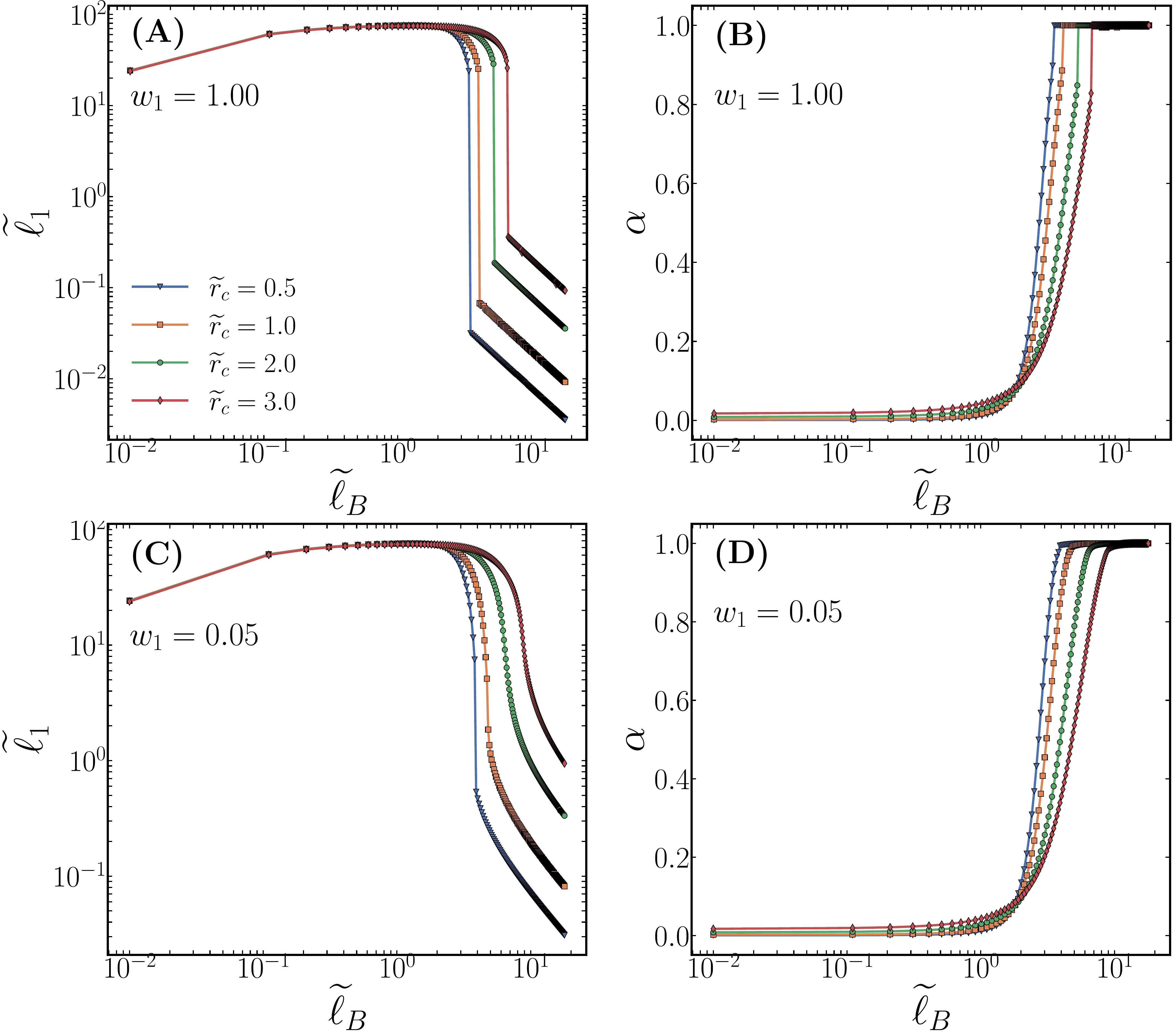}
		\caption{\textbf{Size and charge of a single, isolated, flexible PE chain (including excluded volume and dipolar interactions):} (A) The size $(\widetilde{\ell}_1)$ and (B) the degree of counterion condensation $(\alpha=1-f)$  of the PE chain are plotted as functions of the Bjerrum length $\widetilde{\ell}_B$ for different counterion sizes ($\tilde{r}_c=0.5,1,2,3$), which affect $\delta=\epsilon \ell/\epsilon_l d_{mc}$. Excluded volume and dipolar interactions, assuming freely rotating dipoles, along with three-body interactions are included ($w, w_3$ are calculated). $\epsilon/\epsilon_l$ is fixed at 2.0. The PE chain undergoes first-order coil-globule transition for higher $\widetilde{\ell}_B$, but remains relatively swollen in the collapsed state for bulkier counterions. In (C) and (D) same plots are made with a parametrically reduced dipolar attraction ($w_1=0.05$, instead of 1.0), which show, with bulkier counterions, the PE chain remains relatively swollen, and the transition becomes second-order. The other parameters are: $N=1000,~\bar{c}_s=0.0$, and $\bar{\rho}=0.0005~\tilde{r}_c^3$.}
		
		\label{fig:spm-l1-f-vs-lbt-varying}
	\end{figure}
	\begin{figure}
	\centering
	\includegraphics[width=1\linewidth]{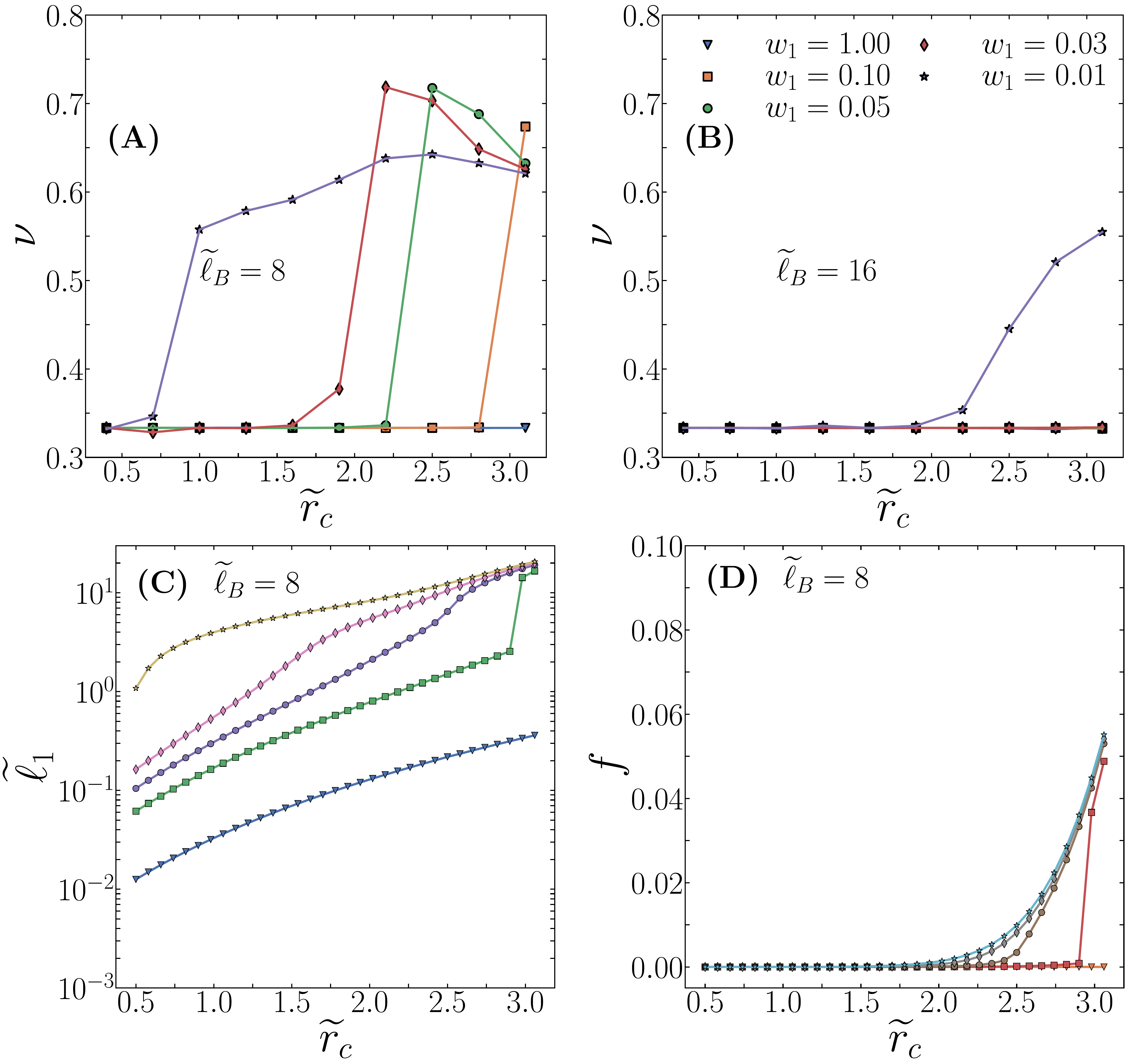}
	\caption{\textbf{Effect of Counterion Size on PE Chain Size and Scaling Exponent at high Coulomb Strengths:} The scaling exponent $\nu$, calculated as $\nu=\log(\widetilde{R}_g)/\log(N)$, is plotted as a function of $\widetilde{r}_c$ for different values of dipolar interaction strength parameter, $w_1$, for two different Coulomb strengths, $\widetilde{\ell}_{B}=8$ (A) and $\widetilde{\ell}_{B}=16$ (B) [lines are guides to the eye]. (C) The size $(\widetilde{\ell}_{1})$ and (D) the degree of ionization $(f)$ are plotted as functions of $\widetilde{r}_c$ for different values of $w_1$, at $\widetilde{\ell}_{B}=8$ and $N=1000$. The scaling exponent indicates the conformational behavior of the PE chain, $1/3$ for a collapsed state due to dipolar attractions of counterion-monomer pairs, $\sim 3/5$ for a swollen state due to excluded volume of bulky counterions, and $\sim 0.7$ for a swollen state due to like-charge repulsions. The chain may remain swollen at high $\widetilde{\ell}_{B}$'s if the counterions are bulkier. The other parameters are: $~\bar{c}_s=0.0$, and $\bar{\rho}=0.0005~\tilde{r}_c^3$.}
	\label{fig:rg-vs-n-lb8-diff-rc}
	\end{figure}

\subsubsection{Effect of counterion size on the conformational behavior of the PE chain}	

	As before, equilibrium values of the chain size $\widetilde{\ell}_{1}$ and the degree of counterion condensation $\alpha$ are plotted as functions of $\widetilde{ \ell}_B$, but this time for different counterion sizes, in Fig.  
 \ref{fig:spm-l1-f-vs-lbt-varying}(A-B). With increasing $\widetilde{ \ell}_B$, counterions condense onto the chain backbone, forming dipoles. The attractive interaction among dipoles, which is taken as a two-body short-range attraction effectively increasing the solvent poorness (Eqs. \ref{F5} and \ref{excluded-volume}), induces a coil-to-globule transition for a sufficiently high electrostatic strength. For example, the  transition occurs for $\widetilde{r}_c=1.0$ at $\widetilde{\ell}_B\sim 4$. The collapse occurs when the dipolar attraction overcomes the electrostatic repulsion among charge uncompensated monomers, resulting in a negligible total charge within the globule due to counterion adsorption that minimizes the electrostatic energy penalty. 
	
	At lower values of $\widetilde{\ell}_{B}$, there is no effect of counterion size on the chain's conformational behavior (the value of $\widetilde{\ell}_{1}$ remains the same up to $\widetilde{\ell}_B\sim1.5$), as shown in Fig. \ref{fig:rg-vs-n-lb8-diff-rc}(A), due to the absence of counterion condensation. Counterion size was found not to impact the conformation of PE chains at low $\widetilde{\ell}_B$ in recent simulations\cite{Gordievskaya2018-SM,Kos2020} too. However, as the electrostatic interactions become more significant, counterions start to condense, and the size of the counterions begins to play a crucial role in determining the chain's behavior. In particular, the chain collapses at a higher value of $\widetilde{\ell}_{B}$ for bulkier counterions as the contribution from the excluded volume increases, as well as there is less overall dipolar attraction for the chain due to less number of dipoles formed. The collapse of the PE chain due to attractive dipole-dipole interactions to a compact globule for smaller counterions has been observed in previous\cite{Winkler1998, Liu2002} and recent\cite{Gordievskaya2018-SM, Gavrilov2016} simulations. Additionally, in Fig. \ref{fig:spm-l1-f-vs-lbt-varying}(B) we note that the PE chain collapses with a slightly lesser degree of condensation for bulkier counterions. As the dipole length is bigger for larger counterions, it results in a stronger dipolar attraction that suppresses the short-range repulsion effects, but it requires a larger $\widetilde{\ell}_{B}$ to do that.
	
		The collapse and related size scaling of the PE chain can be analyzed noting that for $w>0$ one may ignore the $w_3$ term ($F_6$ in Eq. \ref{F6}), take the first derivative of $ F_5 $ with respect to $\widetilde{ \ell}_1$ (Eq. \ref{F5}), and use the definition $ \widetilde{R}_g=\sqrt{{N \tilde{\ell}_{1}}/{6}}$ (from Eq. \ref{rg_def}). For $w<0$, however, $F_6$ needs to be included in the derivative. These lead to
	\begin{align}
		\widetilde{R}_g\sim\begin{cases}
			\left[\left(9/2 \pi \right)^{3 / 2}/\sqrt{6}\right]^{1 / 5}w^{1 / 5} N^{3 / 5}, & w>0 \\
			(2\pi/9)^{1/2}\left(2w_3/w\right)^{1/3} N^{1/3}. & w<0.
		\end{cases}\label{rg-scaling}
	\end{align}
	Substitution of the above equations in the free energy of the PE chain ($F_5+F_6$) in the limits of $w>0$ and $w<0$ gives
	\begin{align}
		\frac{F_5+F_6}{k_B T}\sim\begin{cases}
			(9/2\pi)^{3/5}~w^{2 / 5} N^{1 / 5}, & w>0 \\
			(9/2\pi)^3 (w^2/2w_3)N, & w<0.
		\end{cases}
	\end{align}
	Hence, for $w_{d d}<0$, Eq. \ref{F5} shows that the chain may undergo coil-to-globule transition with respective size scaling exponents, depending on the degree of counterion condensation. This will be validated by results obtained below (in Fig. \ref{fig:rg-vs-n-lb8-diff-rc}).

	Till now we have considered the conformational and counterion adsorption thermodynamics of the PE chain without parameterizing the dipole-dipole interaction. In our theory, we assume freely rotating dipoles, which are progressively more valid at higher temperatures.\cite{Israelachvili} Considering axially restricted dipolar rotations\cite{Cherstvy2010} can lead to significantly reduced attractive interactions, and short-range repulsive potentials may prevent the complete chain collapse in simulations.\cite{Cherstvy2010, Gordievskaya2018-SM} Moreover, due to the presence of a polar solvent, the spatial dielectric behavior may also alter the dipolar interaction strength. In this context, we explore the effects of altered interaction strengths by phenomenologically parameterizing the dipole-dipole interaction using the parameter $w_1$, taking it as a coefficient of the second term in Eq. \ref{finalwdd}. $w_1$, ideally, may have a temperature dependency\cite{Muthukumar2004}. 
	
	With $w_1=0.05$, which corresponds to a significantly reduced dipolar attraction, the  discrete jump in the coil-globule transition is suppressed [Fig. \ref{fig:spm-l1-f-vs-lbt-varying}(C-D)]. For bulky counterions ($\widetilde{r}_{c} > 1$), the chain ceases to undergo a first-
order coil-globule transition, even at high values of $\widetilde{\ell}_{B}
$, and the size reduction is continuous. This behavior may be attributed to weaker dipolar attractions and comparatively strong excluded volume repulsion. In other words, the effective solvent poorness due to dipolar attraction has significantly 
reduced with $w_1=0.05$. As a result, the first-order transition is 
prevented, and the chain remains in a relatively swollen conformation [of 
order Gaussian size, see for $\tilde{r}_c=2.0$ or $ 3.0 $ in 
Fig. \ref{fig:spm-l1-f-vs-lbt-varying}(C)] even at high electrostatic strengths. Similar trends are visible in simiulations\cite{vonFerber2003,Gavrilov2016,Gordievskaya2018-SM}.
	
	To gain further insight into the chain statistics at different  $\widetilde{\ell}_{B}$, we calculated the scaling exponent $\nu$, defined as $\nu=\log(\widetilde{R}_g)/\log(N)$, for different counterion sizes. In Figs. \ref{fig:rg-vs-n-lb8-diff-rc}(A) and \ref{fig:rg-vs-n-lb8-diff-rc}(B) we present the results for $\nu$ as a function of counterion size at $\widetilde{\ell}_{B}=8$ and $\widetilde{\ell}_{B}=16$, respectively, for different values of $ w_1 $. At $\widetilde{\ell}_{B}=8$, $\nu$ is approximately $1/3$ for small counterions, as all of them condense onto the chain, and dipolar attraction collapses the chain to a compact globule. $\nu$ increases to approximately $3/5$ as the counterion size increases, because larger counterions offer higher excluded volume repulsion which eventually overcomes the collapse. With lower values of $w_1$, the chain remains swollen even with counterions of the same size as of the monomers, because dipolar attractions are not strong enough to collapse the chain.
	
	At $\widetilde{\ell}_{B}=16$, the scaling exponent $\nu$ is approximately $1/3$ for all ion sizes except for very low $w_1$ values, because the dipolar attraction becomes strong enough to overcome the excluded volume repulsion even for large counterions. For lower values of $w_1$, such attractions become weak enough to allow excluded volume interactions swell the chain which leads to $\nu=3/5$ for bulkier counterions, as argued in Eq. \ref{rg-scaling}. Within the collapsed globule, where the charge becomes negligible, the interplay between two short-range interactions - dipolar attraction and excluded volume repulsion - becomes crucial. Increasing the counterion size increases the dipole length, and enhances the attractive contributions from the dipole pairs. However, for low values of the strength parameter $w_1$, such gain is limited. The collapse of the chain is then constrained by relatively stronger excluded volume interactions with large counterions, preventing a coil-to-globule transition, whereas such transitions and the collapsed state remain energetically favorable for smaller counterions. Such examples of excluded volume effects competing with electrostatic attractions are available in the literature. Long surfactant tails are found to prevent PE chains from forming collapsed globules\cite{vonFerber2003}. Bulky counterions lead to swelled conformations of PE chains\cite{Gordievskaya2018-SM} and dendrimers\cite{Kos2020} even at high $ \widetilde{\ell}_B $. This behavior contradicts earlier theoretical predictions of collapsed states at high electrostatic regimes\cite{Kundu2014} but aligns with recent simulations\cite{Gordievskaya2018}, which attribute the anomaly to the absence of consideration of steric hindrance in dipole pair interactions.
	
	In Fig. \ref{fig:rg-vs-n-lb8-diff-rc}(C-D), the size ($\widetilde{\ell}_1$) and charge ($f$) of the chain are plotted for different $w_1$ values at $\widetilde{\ell}_{B}=8$, for $N=1000$. Both the size and charge increase with $\widetilde{r}_c$ as mentioned earlier [shown in Fig. \ref{fig:spm-l1-f-vs-lbt-varying}(C-D)]. However, a discontinuous behavior in $\widetilde{\ell}_1$ is observed for $w_1=0.10$ due to the cooperative effects of charge and excluded volume interactions [$f=0.039$ for $\widetilde{r}_c=3$ while $ f=0.0 $ for $\widetilde{r}_c=2.5$, as shown in Fig. \ref{fig:rg-vs-n-lb8-diff-rc}(D)]. Consequently, the scaling exponent $\nu$ changes from $\sim1/3$ to more than 3/5 with $\widetilde{r}_c$ going from 2.5 to 3.0 [Fig. \ref{fig:rg-vs-n-lb8-diff-rc}(A)]. Note that the scaling parameter is higher than $ 3/5 $, due to the presence of like charge repulsion from uncompensated monomers, as seen before in simulations.\cite{Winkler1998}
	
	In Fig. \ref{fig:rg-vs-n-lb8-diff-rc}(A) too, we note that the scaling exponent rises to $\sim 0.70$, for example in the case of $w_1=0.05$ for which $f=0.01$ for $\widetilde{r}_c=2.5$, due to the like charge repulsion of charge uncompensated monomers. In some simulations with highly charged PEs, the scaling exponent is found to reach as high as ~$ 1.0 $ and collapse to $ 0.33 $.\cite{Winkler1998,Chi2013}. However, in Fig. \ref{fig:rg-vs-n-lb8-diff-rc}(A), the exponent comes down to $\sim 3/5$ for $\widetilde{r}_c > 2.5$, due to increased excluded volume interactions from condensed bulkier ions (note that $f=0.04$, which is not zero and even higher for $\widetilde{r}_c=3.0$). The reversal of the trend in the exponent $\nu$ with ionic size can be attributed to the relative influences of like charge repulsion of uncompensated monomers and excluded volume repulsion of the condensed counterions.
	
	Note that for $w_1=1$, in collapsed states [Fig. \ref{fig:spm-l1-f-vs-lbt-varying}(A), (C)], the value of $\widetilde{\ell}_{1}$ increases with $\widetilde{r}_c$. The competitive effects of the two-body and three-body interactions [the second term of Eq. \ref{F5} and Eq. \ref{F6}, and also as argued in Eq. \ref{rg-scaling}] keeps the polyelectrolyte size larger for bulkier counterions. Despite the size variation of the PE chain due to the presence of counterions of variable sizes, the size scaling exponent remains the same ($\nu\sim1/3$) for all collapsed cases, as expected from Eqs. \ref{excluded-volume} and \ref{rg-scaling} [shown in Fig. \ref{fig:rg-vs-n-lb8-diff-rc}(B)].\cite{Gavrilov2016}
		
	\subsubsection{Thermodynamics: effect of counterion size on the entropy-enthalpy interplay in counterion condensation}\label{FE-components}
		\begin{figure}[!htbp]%
		\centering
		\includegraphics[width=0.49\linewidth]{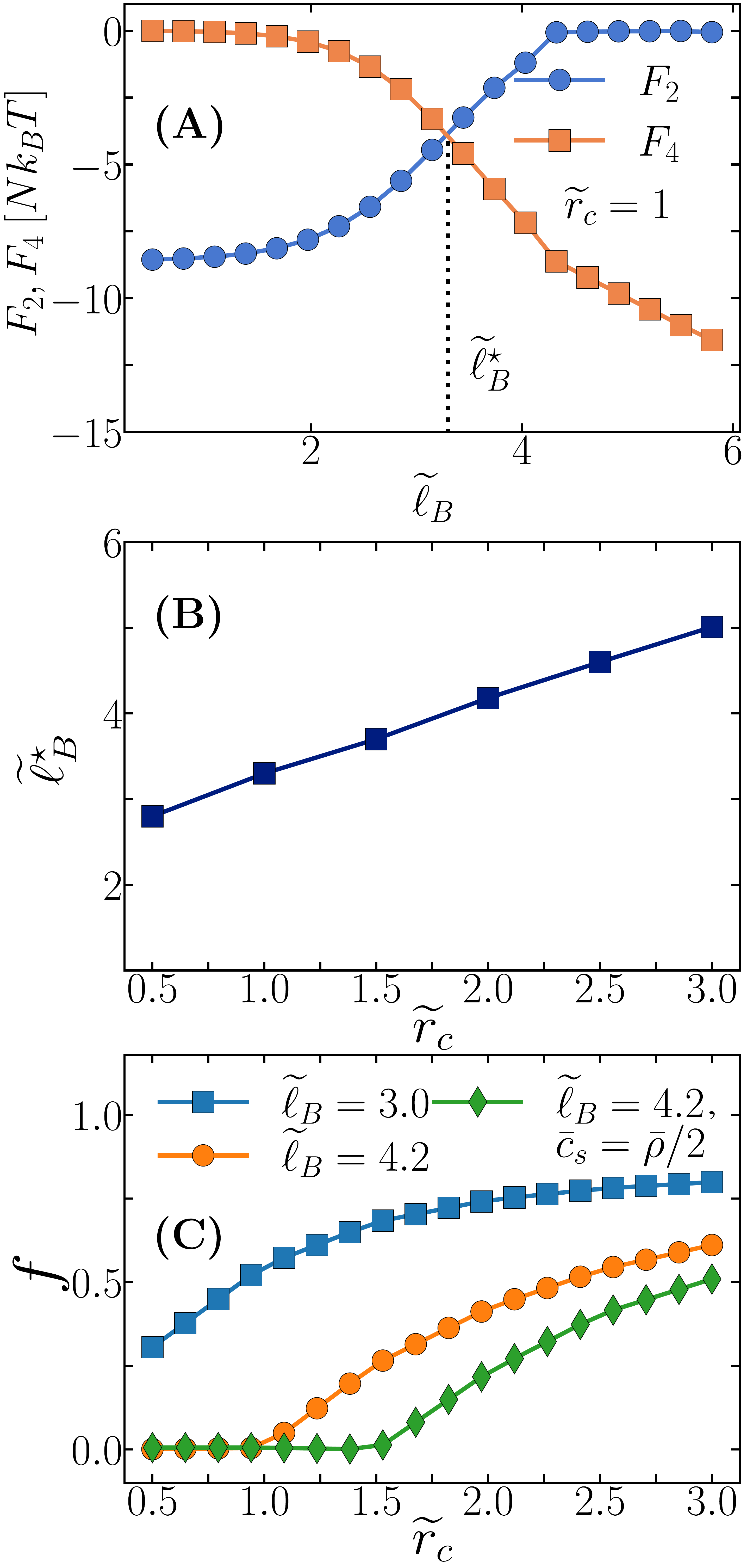}
		\includegraphics[width=0.49\linewidth]{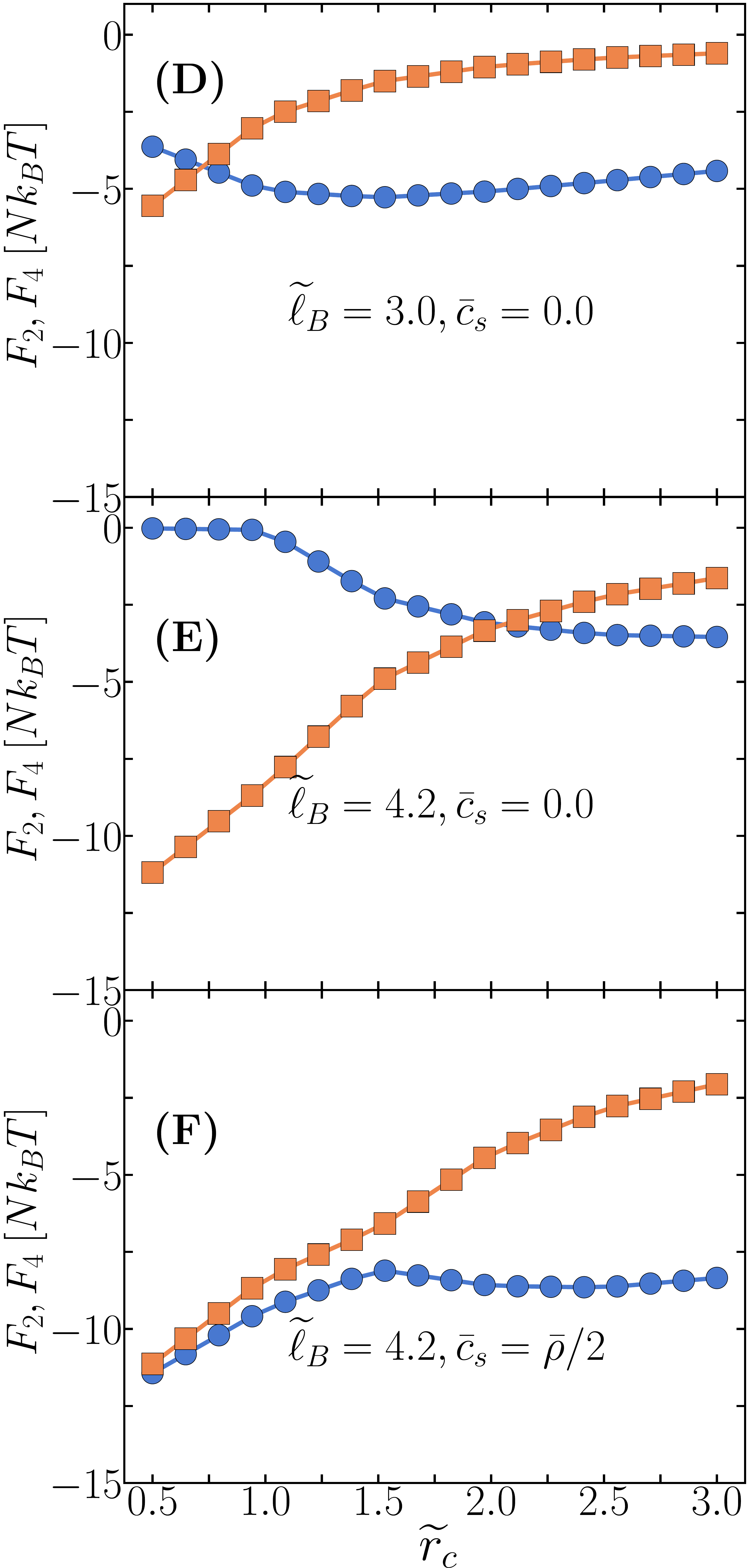}
		\caption{\textbf{Effect of Counterion Size on Thermodynamics and Interplay of Free Energy Components:} 
			(A) The variation of entropic free energy of free ions ($F_2$) and electrostatic free energy of counterion-monomer pair formation ($F_4$), in units of $ N k_{B}T $, are plotted as functions of $\widetilde{\ell}_{B}$ (keeping $\tilde{r}_{c}=1$ and $\bar{c}_{s}=0.0$). $F_4$ gains at the loss of $F_2$ with increasing $\widetilde{\ell}_{B}$. (B) The crossover value, $\widetilde{\ell}_{B}^{\star}$, where $F_2=F_4$, increases with $\tilde{r}_{c}$ (salt, $\bar{c}_{s}=0.0$), indicating that smaller counterions condense more, which is confirmed with (C) the ionization degree $f$ increasing with $\tilde{r}_{c}$ for different values of $\widetilde{ \ell}_{B}$ and $ \bar{c}_{s} $. Higher salt induced more condensation. Panels (D)-(F) exhibit $F_2$ and $F_4$, in units of $ N k_{B}T $, as functions of $\tilde{r}_{c}$ for different values of $\widetilde{\ell}_{B}$ and $\bar{c}_s$. Specifically, panel (D) shows the results for $\widetilde{\ell}_{B}=3.0$ and $\bar{c}_s=0.0$, (E) for $\widetilde{\ell}_{B}=4.2$ and $\bar{c}_s=0.0$, and (F) for $\widetilde{\ell}_{B}=4.2$ and $\bar{c}_s=\bar{\rho}/2$. Non-monotonic trends of $F_2$ are seen due to competition between entropy and enthalpy components. The other parameters are: $ N=1000, ~\bar{\rho}=0.0005~\tilde{r}_c^3, ~\epsilon/\epsilon_l=2$.}
		\label{fig:spm-f2-f4-vs-lbt-lbstar-vs-rc}
	\end{figure}

	As discussed before, the PE chain collects counterions from the solution, primarily driven by the competing free energy contributions from the translational entropy of the free ions ($F_2$) and ion-pair formation ($F_4$). Both contributions decrease (which is a gain in free energy) with a decreasing counterion size. Therefore, it is not straightforward to predict the outcome of a change in size of counterions intuitively. However, for most results studied in this work, and for the set of modest parameter values used, a smaller counterion size is found to result in a higher degree of counterion condensation. This implies that the ion-pair free energy gain ovecomes the entropic loss due to the resultant depletion of free ions. A part of the ion-pair free energy can be entropic, one may note, due to reorganization of solvent dipoles,\cite{Chen2022-PNAS} but how significant that part is for a PE chain with organic backbone and fractal geometry is a matter of discussion.\cite{Mitra2023,Ghosh2023} We consider this free energy ($F_4$) enthalpic in this work, although it must be taken a nominal quantity\cite{Chowdhury2023}.

	To analyze this interplay of enthalpy and entropy, in Fig. \ref{fig:spm-f2-f4-vs-lbt-lbstar-vs-rc}, we look at the thermodynamics by plotting $F_2$ and $F_4$, in units of $N k_B T$, as functions of the Bjerrum length $\widetilde{\ell}_{B}$ (proportional to $1/\epsilon T$). First, in Fig. \ref{fig:spm-f2-f4-vs-lbt-lbstar-vs-rc}(A), we take the counterions of the same size as the monomers ($\widetilde{r}_c=1$). At low $\widetilde{\ell}_{B}$, the PE is unable to collect counterions from the solution due to weak electrostatic correlations compared to thermal fluctuations. Consequently, $F_2$ is higher than $F_4$. As 
$\widetilde{\ell}_{B}$ increases, more counterions are attracted from the solution at the cost of their translational entropy. This leads to an energy gain with a decreased $F_4$ and increased counterion condensation, reducing the number of counterions in the solution as well as $F_2$ [also see Fig. \ref{fig:spm-l1-f-vs-lbt-fixed}(B) and Fig. \ref{fig:spm-l1-f-vs-lbt-varying}(B), (D)]. 
	
	At an intermediate value of $\widetilde{\ell}_{B}$, which we define as a crossover point denoted as $\widetilde{\ell}_{B}^{\star}$, $F_2$ becomes equal to $F_4$. The crossover parameter $\widetilde{\ell}_{B}^{\star}$ is found to increase monotonically with the ionic size, as depicted in Fig. \ref{fig:spm-f2-f4-vs-lbt-lbstar-vs-rc}(B). Here too, bulkier counterions condense less with less gain in electrostatic energy, that requires a higher $\widetilde{\ell}_{B}^{\star}$ to win over the translational entropy (although, note again, that the translational entropy is also less for larger counterions).  This leads to the monotonic increase of $\widetilde{\ell}_{B}^{\star}$ with $\widetilde{r}_c$. The state boundary we defined with $\widetilde{\ell}_{B}^\star$ is a unique line where the charge of the chain (degree of counterion condensation) is found to be a constant quantity irrespective of the counterion size (results not shown). 
	
	In addition to the effects of counterion size and the Bjerrum length, introduction of salt, expectedly, induces more counterion condensation (higher $\alpha$), or a reduction in the degree of ionization, $f$ [Fig. \ref{fig:spm-f2-f4-vs-lbt-lbstar-vs-rc}(C)]. $ F_2 $ and $ F_4 $ are plotted for two values of $\widetilde{\ell}_{B}=3$ and $4.2$, in Fig. \ref{fig:spm-f2-f4-vs-lbt-lbstar-vs-rc}(D) and (E), respectively. The chosen values of $\widetilde{\ell}_{B}$ are such that one is less than $\widetilde{\ell}_{B}^{\star}$ and the other corresponds to the minimum value of $\widetilde{\ell}_{B}$ where all counterions of the size of the monomer are condensed.

	As smaller counterions condense more [Fig. \ref{fig:spm-f2-f4-vs-lbt-lbstar-vs-rc}(C)], fewer remain to roam free in the bulk solution, contributing  less entropically to the free energy [Fig. \ref{fig:spm-f2-f4-vs-lbt-lbstar-vs-rc}(D)]. This has been experimentally observed in PE gels, having decreasing conductivity with small counterions indicating a decrease in free ion concentration\cite{Rumyantsev2016}. It is indeed a counterintuitive result that the bulkier counterions being less condensed have more translational entropy collectively [Fig. \ref{fig:spm-f2-f4-vs-lbt-lbstar-vs-rc}(D)]. However, once most of the counterions become free with increasing counterion size ($\tilde{r}_c \sim 2$), then with further increase of size the translational entropy decreases, albeit slightly, as expected [Fig. \ref{fig:spm-f2-f4-vs-lbt-lbstar-vs-rc}(D)]. For a high electrostatic strength (here, $\widetilde{\ell}_{B}=4.2$), all smaller counterions are condensed [Fig. \ref{fig:spm-f2-f4-vs-lbt-lbstar-vs-rc}(C)], and $ F_2 $ approaches zero [Fig. \ref{fig:spm-f2-f4-vs-lbt-lbstar-vs-rc}(E)]. However, at that $\widetilde{\ell}_{B}$, not all bulkier counterions are condensed, leading to an increase in the entropic contribution to the free energy ($ F_2 $) from the free ions as their size increases, ultimately surpassing the enthalpic contribution from ion-pair formation ($ F_4 $). Note that the crossover value of $r_c$ for $\widetilde{\ell}_{B} = 4.2$ is higher than that of $\widetilde{\ell}_{B} = 3.0$. This is due to the increased electrostatic energy gain with smaller ions. Therefore, decreasing $r_c$ and increasing $\widetilde{\ell}_{B}$ have similar thermodynamic effects on the system, as shown in Fig. \ref{fig:spm-f2-f4-vs-lbt-lbstar-vs-rc}(D-E).
	
	In the case of added salt, the counterion adsorption is enhanced [Fig. \ref{fig:spm-f2-f4-vs-lbt-lbstar-vs-rc}(C)]. It results in a change in the crossover value of $\widetilde{\ell}_{B}$. The gain in the electrostatic energy of the ion-pairs ($F_4$) remains similar (but slightly increased with slightly more counterion condensation), but the presence of salt provides significant entropic free energy (negative $F_2$) even for smaller counterions, and even when all monomeric charges are compensated by condensed counterions [Fig. \ref{fig:spm-f2-f4-vs-lbt-lbstar-vs-rc}(F)]. Such a contribution diminishes with an increased size of counterions, but only to some extent. $F_2$ exhibits such a non-monotonic dependence on $r_c$, due to the interplay between electrostatics and excluded volume effects. Weaker electrostatic correlations cause more bulky counterions to stay in the solution, resulting in a gain in entropy. However, due to the ionic size the available volume reduces, leading to a non-monotonic behavior in $F_2$ [Fig. \ref{fig:spm-f2-f4-vs-lbt-lbstar-vs-rc}(D), (F)]. The non-linearity and unpredictability of the thermodynamics in the presence of salt, for a varying size of counterions, is truly manifest in Fig. \ref{fig:spm-f2-f4-vs-lbt-lbstar-vs-rc}(F).
	
	\section{summary}
		
	We present a theory to investigate the influence of counterion size on the effective charge, size, and thermodynamics of a single, isolated, and flexible polyelectrolyte chain. Our analysis takes into account the effects of counterion size on various factors, including the ion-pair energy, excluded volume effect, volume entropy of free ions, volume entropy of condensed ions, dipolar interactions captured through the second virial coefficient, and the third virial coefficient. The increase in ion size reduces the gain in free energy due to both ion pair formation and the entropy of free ions. As a result, it leads to non-monotonic effects in the system, making it difficult to predict the consequences of variable ion sizes intuitively. 
	
	In the model, we apply the Edwards-Muthukumar interaction Hamiltonian, which captures the self-energy of the PE chain through segment–segment electrostatic and excluded volume interactions, including dipolar interactions, and also the conformational entropy of the PE chain. We assume a Gaussian monomer density profile, following Flory, which simplifies the analysis. We consider the finite size of counterions in calculating the entropy of condensed counterions assumed confined within a cylindrical volume surrounding the PE chain backbone. Importantly, we incorporate counterion size effects in the dipole pair interactions, which include short-range repulsions in addition to the dipolar attractions.  Minimization of the total free energy, that treats the Bjerrum length, dielectric mismatch, and counterion size as parameters, determines the PE's effective charge and size. Our primary focus has been to understand conformational and thermodynamic aspects in the presence of finite-size counterions, offering a direct means to evaluate how experimental factors like electrostatic strength and salt influence chain conformations, PE chain scaling exponent, and thermodynamics.

	We benchmarked the results against previous established studies and found that the size of the PE chain varies non-monotonically with the electrostatic strength $\delta \widetilde{\ell}_{B}$, the product of dielectric mismatch and Bjerrum length, as expected. At modest electrostatic strengths, the chain is swollen due to like-charge repulsion, but it forms a globule at higher strengths due to counterion condensation. The attractive interaction of the freely rotating dipoles leads to an abrupt, first order coil-to-globule transition, which occurs when the short-range dipolar attraction overcomes both the electrostatic repulsion between monomers and excluded volume effects. This dramatic collapse has been absent in simulations where it is found to be a continous, second order transition. Considering a potential overestimation of the dipolar interactions (because freely rotating dipoles are valid at high temperatures and in polar solvents), we introduced a phenomenological strength parameter $ w_1 $ to moderate the dipolar attraction. Reduced values of $ w_1 $ lead to continuous transitions as expected. The size scaling exponent $\nu$ is found to be approximately $1/3$ for small counterions, as the chain is collapsed to a globule with inadequate excluded volume repulsion. However, $\nu$ increases to $3/5$ for large counterions with increased excluded volume repulsion. If electrostatic repulsion between charge-uncompensated monomers is present, the exponent increases to values greater than 3/5, up to 0.70 in some cases. In addition, we also analyzed the thermodynamic interplay between the free ion entropy ($F_2$) and ion-pair formation energy ($F_4$), considering the latter to be nominal ignoring contributions from orientation of solvent dipoles. The point of crossover where $F_2$ equals $F_4$, denoted as $\widetilde{\ell}_{B}^{\star}$, is found to monotonically increase with ionic size. Larger counterions tend to stay more in the bulk solution due to weaker electrostatic attraction. But due to their larger ionic size, the available volume also reduces, so does the entropy, and it leads to the non-monotonic behavior of $F_2$. 

\section{Acknowledgment}
	
The authors acknowledge financial support from IISER
Kolkata, Ministry of Education, Government of India. They also
thank Soumik Mitra, Aritra Chowdhury, and Benjamin Schuler for discussions and other collaborative work which helped better understand the role of counterions in polyelectrolyte systems.

	\bibliography{2023-10-17-SG-AK-arxive-Submission}
\end{document}